\documentclass[10pt,letterpaper]{article}
\usepackage{opex3}

\begin{document}

%%%%%%%%%%%%%%%%%% title page information %%%%%%%%%%%%%%%%%%
\title{Observation of electromagnetically induced transparency for a squeezed vacuum with the time domain method}

\author{M. Arikawa,$^1$ K. Honda,$^{1,2}$ D. Akamatsu,$^1$ Y. Yokoi$^1$, \\K. Akiba,$^1$ S. Nagatsuka,$^1$  A. Furusawa,$^3$ and M. Kozuma$^{1,2,4}$}

\address{$^1$Department of Physics, Tokyo Institute of Technology, \\ 2-12-1 O-okayama, Meguro-ku, Tokyo 152-8550, Japan}
\address{$^2$Interactive Research Center of Science, Tokyo Institute of Technology, \\ 2-12-1 O-okayama, Meguro-ku, Tokyo 152-8550, Japan}
\address{$^3$Department of Applied Physics, School of Engineering, \\ The University of Tokyo, 7-3-1 Hongo, Bunkyo-ku, Tokyo 113-8656, Japan}
\address{$^4$CREST, Japan Science and Technology Agency, 1-9-9 Yaesu, Chuo-ku, Tokyo 103-0028, Japan}

\email{kozuma@ap.titech.ac.jp} %% email address is required

% \homepage{http:...} %% author's URL, if desired

%%%%%%%%%%%%%%%%%%% abstract and OCIS codes %%%%%%%%%%%%%%%%
%% [use \begin{abstract*}...\end{abstract*} if exempt from copyright]

\begin{abstract}
A probe light in a squeezed vacuum state was injected into cold $^{87} $Rb atoms with an intense control light in a coherent state. A sub-MHz window was created due to electromagnetically induced transparency, and the incident squeezed vacuum could pass through the cold atoms without optical loss, as was successfully monitored using a time-domain homodyne method. 
\end{abstract}

\ocis{(270.0270) Quantum optics; (270.6570) Squeezed states.} % REPLACE WITH CORRECT OCIS CODES FOR YOUR ARTICLE

%%%%%%%%%%%%%%%%%%%%%%% References %%%%%%%%%%%%%%%%%%%%%%%%%

%%%%%%%%%%%%%%%%%%%%%%%%%%  body  %%%%%%%%%%%%%%%%%%%%%%%%%%
\section{Introduction}
 The coherent transfer of quantum information between light and atoms has been actively investigated, and various methods for its implementation using several phenomena, including the Raman process \cite{RamanProcess}, quantum non-demolition measurement \cite{QND}, and electromagnetically induced transparency (EIT) \cite{DarkSP,EIT}, have been proposed and investigated experimentally. Experimental demonstration of storage and retrieval of a single photon state was first realized using EIT \cite{SinglephotonKuzmich,SinglephotonLukin}, and this method has attracted a great deal of attention for application to various non-classical lights \cite{StorageParametric}. The  squeezed vacuum has been an important resource for deterministic intricate quantum information processing. Storing and retrieving the squeezed vacuum state thus enables us to apply such quantum information protocols to the spatially localized atomic ensemble. Furthermore, storage of the squeezed vacuum implies atomic spins being squeezed under the standard quantum limit \cite{SSS}, which is useful for precise measurements such as magnetometry \cite{Magnetometry}. The storage and retrieval of light with EIT is based on ultraslow propagation of light by steep dispersion within the transparency window. The group velocity of an incident light pulse can be dramatically reduced by simply narrowing the transparency window. Thus, the establishment of a technique for detecting the squeezed vacuum passed through a narrow transparency window forms the foundation for storage and retrieval of the squeezed vacuum.

The first experimental confirmation of EIT with a squeezed vacuum \cite{EITwithSV} employed the homodyne detection method with a monochromatic local oscillator (LO), where the homodyne detection signal was analyzed using a spectrum analyzer. Squeezing was observed when the frequency width of the transparency window was relatively large (2.6 $\mathrm{MHz}$). It is noted that the spectrum analyzer measures the power spectrum by mixing the input signal with an electric local oscillator (eLO) and frequency filtering the mixer output (IF). Practically, a tiny amount of the eLO directly couples to the IF port of the mixer and this leakage is detected when the eLO frequency is within the filter bandwidth. Eventually, a sharp peak appears at the low frequency region reflecting the 'shape of the filter', which mainly limited the detectable minimum frequency of the squeezing.

In order to solve this problem and realize ultra-slow propagation of a squeezed vacuum, homodyne detection with bichromatic LO having the frequency components of $\nu_0\pm\epsilon$ was utilized \cite{Slow}, where $\nu_0$ and $\epsilon$ were the carrier frequency of the squeezed vacuum and the center frequency of the spectrum analyzer, respectively. Since this method is sensitive to the degenerate frequency component ($\nu_0$), squeezing was observed within the sub-$\mathrm{MHz}$ transparency window and the ultra-slow propagation of a squeezed vacuum pulse was realized. However, the bichromatic homodyne method is also sensitive to the two-mode quadrature noise consisting of $\nu_0 \pm 2 \epsilon$, which are usually present outside the transparency window, since $\epsilon$ has to be set to a relatively high frequency to avoid the specific peak reflecting the 'shape of the filter' of the spectrum analyzer. Thus, the observable squeezing level inevitably decreases in the storage and retrieval process.

Most recently, squeezing was successfully observed for the low-frequency region by Fourier transform of the real-time homodyne detection signal \cite{Odd,TimeGated}. In this Letter, we report the successful observation of a squeezed vacuum that has passed through a sub-$\mathrm{MHz}$ EIT window using the time domain method. The homodyne detection with the spectrum analyzer provides the power noise of the quadrature amplitude. In contrast, the time domain method acquires real-time signal of the quadrature amplitude and thus we can extract frequency spectrum of the quadrature noise by simply performing Fourier analysis. It should be noted that the previously observed narrow band EIT spectrum \cite{Slow} does not give us direct information about how the squeezed vacuum is degraded by the EIT medium under the slow propagation condition. In the previous experiment, the spectrum was obtained by measuring the squeezing level for various detuning of the control light, while the slow propagation was carried out for the fixed control frequency. The FFT analysis enables us to obtain the squeezing spectrum for the fixed control frequency, which is direct information about the response of the squeezed vacuum for the EIT medium under the condition utilized for the ultra-slow propagation and storage experiments.

\section{Experiment}
 Our experimental setup is shown schematically in Fig. 1. We prepared magneto-optically trapped and laser cooled $^{87}$Rb atoms and employed them as the EIT medium. One cycle of our experiment was 10 ms. Each cycle consisted of the preparation of cold atoms (9 ms) and the measurement of EIT (1 ms). During the preparation period, the Rb gas was magneto-optically trapped for 5.5 ms, after which the magnetic field was turned off. After the eddy current ceased ($\sim$3 ms), both the cooling and repumping lights were turned off and depumping light tuned to F=2 $\to$ F'=2 transition was made incident on the gas to prepare cold atoms in the F = 1 state.
 
\begin{figure}[htbp]
\centerline{\includegraphics[width=11cm]{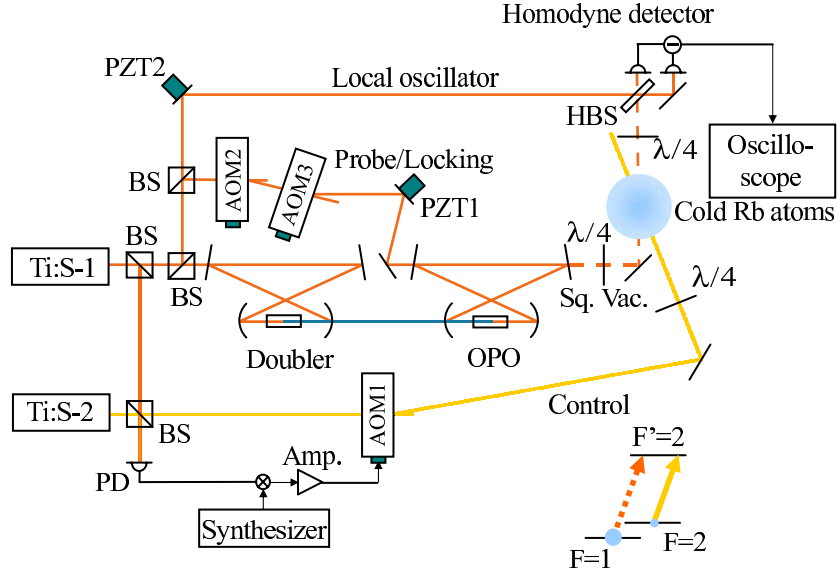}}
\caption{Schematic diagram of the experimental setup. BS: beam splitter, HBS: half beam splitter, AOM: acousto-optic modulator, PD: photodetector, PZT: piezo electric transducer, Amp.: RF amplifier, Sq. Vac.: squeezed vacuum.}
    \label{fig:1.eps}
\end{figure}
  
 We first evaluated the frequency width of the EIT window with a weak probe light in a coherent state. Ti:Sapphire laser 1 was tuned to 5$^{2}$S$_{1/2}$, F=1 $\to$ 5$^{2}$P$_{1/2}$, F'=2, which corresponds to a probe transition. A weak light from this laser was made incident on the OPO cavity, and the output beam was used as a probe light. Note that the probe light was in a coherent state because the second harmonic light in front of the OPO cavity was blocked. The procedure described above enabled us to employ a coherent state of the probe light with a spatial mode that was identical to that of the squeezed vacuum used in the later experiment. Ti:Sapphire laser 2 was tuned to 5$^{2}$S$_{1/2}$, F=2 $\to$ 5$^{2}$P$_{1/2}$, F'=2 and was used for the control light. The relative frequency between the probe and the control lights was stabilized with a synthesizer and AOM1 using a feed-forward method \cite{Slow, FeedFoward}. During the measurement period, the probe and the control lights were incident on the cold Rb gas with a crossing angle of 2.5$^\circ$. The power of the control light was approximately 80 $\mu$W. The waists of the probe and the control lights were 150 $\mu$m and 550 $\mu$m, respectively. These lights were circularly polarized in the same direction by using quarter waveplates ($\lambda/4$). It is noted that the atomic sample prepared here was hyperfine pumped, but unpolarized. Choosing the same circular polarizations for both the probe and the control lights allowed us to keep high transparency while the Zeeman degeneracy of the atomic levels was concerned \cite{SinglephotonKuzmich}.The probe light that passed through the cold atoms was photon-counted using a silicon avalanche photodiode (Perkin-Elmer: SPCM-AQR). The transmission spectrum obtained by scanning the frequency of the control light is shown in Fig. 2, where the atoms were almost transparent around two-photon resonance, and the half width at half maximum was approximately 300 $\mathrm{kHz}$. The slight asymmetry of the signal originated from the atomic density variation during the measurement time due to thermal diffusion.

\begin{figure}[htbp]
\centerline{\includegraphics[width=6cm]{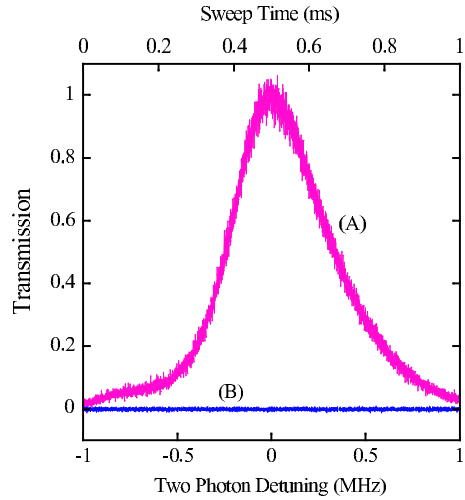}}
\caption{ Transmission spectrum of the coherent probe light as a function of the two-photon detuning. Trace (A) indicates transmission spectrum when the probe light was incident on the cold atoms with the control light. Trace (B) shows the spectrum without the control light.}
    \label{fig:2.eps}
\end{figure}  

 We next observed the EIT spectrum for the squeezed vacuum, which passed through the cold atoms under the EIT condition. We generated the squeezed vacuum resonant on the Rb $D_1$ line using a sub-threshold optical parametric oscillator (OPO) with a periodically poled KTiOPO$_4$ crystal \cite{PPKTP}. Injecting 85 mW of the second-harmonic light from the doubler into the OPO cavity, the squeezed vacuum was generated and was incident on the cold atoms during the measurement period. The relative phase between the squeezed vacuum and the LO had to be fixed in order to selectively monitor anti-squeezing and squeezing.  We stabilized the relative phase with the help of the weak coherent state of light, which was used as the probe in the above classical EIT experiment. We hereinafter refer to this light the locking beam. The relative phase between the locking beam and the second harmonic light was stabilized using a classical parametric amplification signal and piezo transducer PZT1. The relative phase between the locking beam and the LO was also stabilized using the homodyne detector signal and PZT2. Eventually, the relative phase between the squeezed vacuum and the LO was locked to $\theta = \pi/2$ or $0$ during the cold atom preparation period. The feedback voltages to PZTs were held in the measurement period and the weak locking beam was turned off with AOM2 and AOM3, so that we could measure anti-squeezing or squeezing in the measurement period \cite{TimeGated,LockFreeze}.
 
 During the measurement period, we imported signals from the homodyne detector into the high-speed digital oscilloscope and performed Fourier transform of the obtained real-time waveforms, where the signal sampling rate was 5 $\times$ 10$^{7}$ samples/sec. Vertical resolution of the digital oscilloscope was 8 bits and thus the dynamic range for the power measurement was almost 50dB. In order to minimize the attribution of the classical noises to the homodyne signal, we adjusted the power balance between the two LO beams as much as possible. We applied classical power modulation to the LO and minimized the modulation signal from the balanced homodyne detector by changing the reflectivity of the half beam splitter, which was carried out by adjusting the angle of the splitter. Eventually we could achieve $-58$~dB of suppression for the classical modulation. Figure 3(a) indicates the frequency spectrum of the quadrature noise for the squeezed vacuum, where trace (A) indicates the shot noise and traces (B) and (C) indicate the quadrature noises without the cold atoms. Here, the relative phases were set to $\theta = \pi/2$ and $0$, respectively. Each data was averaged over 1,000 times. The sharp peaks in Fig. 3(a) originated from the electric circuit noises of the homodyne detector and thus they could not be eliminated by balancing the laser intensities. The quadrature noise was approximately flat over the frequency range shown in Fig. 3(a) because the spectrum width of the OPO (15 $\mathrm{MHz}$) was much broader than the measurement frequency range (1.5 MHz). Traces (D) and (E) indicate the measured quadrature noise of the squeezed vacuum passed through the cold atoms with the control light, where the relative phase was set to $\theta = \pi/2$ and $0$, respectively. Both anti-squeezing and squeezing were observed within the sub-MHz narrow frequency region, which corresponds to the transparency window caused by EIT.
 
 \begin{figure}[htbp]
\centerline{\includegraphics[width=11cm]{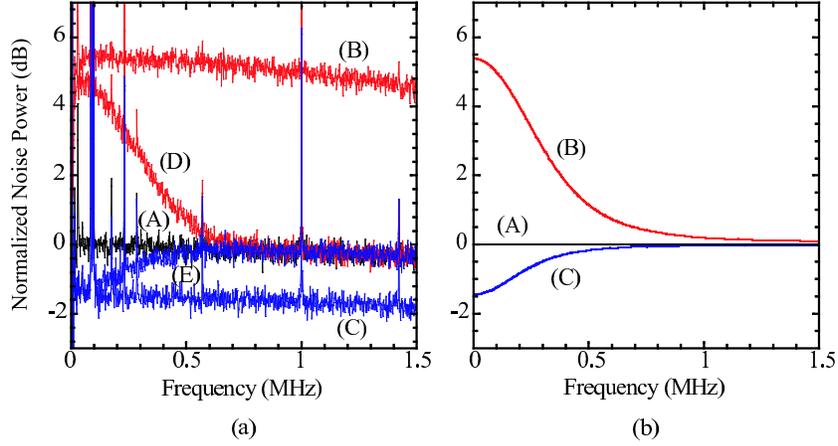}}
\caption{(a) Quadrature noise of the probe light in the squeezed state. Trace (A) indicates the shot noise. Traces (B) and (C) show the quadrature noises of the probe light without cold atoms, where the relative phase were $\theta = \pi/2$ and $0$, respectively. Traces (D) and (E) show the quadrature noises when the probe light was incident on the cold atoms with the control light, where the relative phase were $\theta = \pi/2$ and $0$, respectively. (b) The numerically simulated noise spectrum for the squeezed vacuum passed through the cold atoms under the EIT condition. 
Trace (A) indicates the shot noise. Traces (B) and (C) show the quadrature noises for $\theta = \pi/2$ and $0$, respectively.} 
    \label{fig:3.eps}
\end{figure}

 In a previous experiment \cite{Slow}, the squeezing and the anti-squeezing signals alternatively appeared when the frequency of the control light was swept within a narrow transparency window, which was due to atoms adding an additional phase shift to the squeezed vacuum. Such a complicated structure did not appear in the present experiment as a result of the cancellation of the dispersion effect. Note that in the current experiment, the frequency of the control light was not swept, but rather was fixed, and thus the sidebands constructing the squeezed vacuum always have the opposite sign of the additional phase shifts. The two mode quadrature operator is defined as
\begin{equation} 
 \hat{X}(\nu,\theta)=\hat{a}_{\nu_0+\nu}\exp{(-i\theta)}+\hat{a}^\dag_{\nu_0-\nu}\exp{(i\theta)},
 \label{eq:Xoperator1}
\end{equation}
where the sideband frequencies are $\nu_0 \pm \nu$ \cite{Quadrature}. Now we consider the situation where the opposite sign of the phase shifts $\pm \phi$ are added to the two sidebands,i.e., 
\begin{equation} 
 \hat{X}(\nu,\theta,\phi)=\hat{a}_{\nu_0+\nu}\exp{(-i\theta)}\exp{(i\phi)}+\hat{a}^\dag_{\nu_0-\nu}\exp{(i\theta)}\exp{(-i(-\phi))}.
 \label{eq:Xoperator2}
\end{equation}
The quadrature noise is given by
\begin{eqnarray}
&\langle|\Delta\hat{X}(\nu,\theta,\phi)|^{2}\rangle=\langle(\hat{X}(\nu,\theta,\phi)-\langle\hat{X}(\nu,\theta,\phi)\rangle)(\hat{X}^\dag(\nu,\theta,\phi)-\langle\hat{X}^\dag(\nu,\theta,\phi)\rangle)\rangle_{Sym} \nonumber \\ 
&=\frac{1}{2}\langle\hat{X}(\nu,\theta,\phi)\hat{X}^\dag(\nu,\theta,\phi)+\hat{X}^\dag(\nu,\theta,\phi)\hat{X}(\nu,\theta,\phi)\rangle-\langle\hat{X}(\nu,\theta,\phi)\rangle\langle\hat{X}^\dag(\nu,\theta,\phi)\rangle \nonumber \\
&=\frac{1}{2}\langle\hat{X}(\nu,\theta)\hat{X}^\dag(\nu,\theta)+\hat{X}^\dag(\nu,\theta)\hat{X}(\nu,\theta)\rangle-\langle\hat{X}(\nu,\theta)\rangle\langle\hat{X}^\dag(\nu,\theta)\rangle \nonumber \\
&=\langle|\Delta\hat{X}(\nu,\theta)|^{2}\rangle
 \label{eq:Xoperator3}
\end{eqnarray}
where
\begin{equation}
 \langle\hat{A}\hat{B}\rangle_{Sym}=\langle\hat{A}\hat{B}+\hat{B}\hat{A}\rangle/2.
 \label{eq:Xoperator4}
\end{equation}
Namely, additional phase $\phi$ does not affect the power noise of the quadrature amplitude.

Figure 3(b) shows the numerical simulation of the quadrature noises of the squeezed vacuum that passed through the cold atoms under the EIT condition, where, for the sake of simplicity, we assumed the incident squeezed vacuum experienced the optical loss corresponding to Fig. 2 and the incident anti-squeezing and squeezing levels were constant in all frequency regions. The absorption loss is represented by the model using a beam splitter whose transmittance T($\nu$) is dependent on the frequency of the light. The quadrature noise of the squeezed vacuum experienced absorption loss is given by
\begin{equation}
\langle|\Delta\hat{X}(\nu,\theta)|^{2}\rangle=\frac{1}{4}\{T(\nu)(\cosh 2r-\cos 2\theta \sinh 2r)+(1-T(\nu))\},
\end{equation}
where $r$ is the squeezing parameter. We use the average of the transmission of the coherent probe light passing through the atoms with the control light whose detuning was $\pm \nu$, which is shown in Fig. 2, as T($\nu$).
The numerical simulation is in good agreement with the experimentally obtained noise spectrum.

\section{Conclusion}
 In conclusion, we have succeeded in observing the EIT spectrum of the squeezed vacuum, where both squeezing and anti-squeezing were monitored within sub-MHz transparency window. The observation of squeezing for such a low-frequency region is an essential step in the realization of the storage and retrieval of the squeezed vacuum.

\section*{Acknowledgment}
 The authors would like to thank N. Takei for engaging in numerous helpful discussions. Two of the authors (D. A. and K. A.) were supported 
by the Japan Society for the Promotion of Science. This study was supported by a Grant-in-Aid for Scientific Research (B) and the 21st Century COE Program at Tokyo Tech, ''Nanometer-Scale Quantum Physics'' by the MEXT.
 

\begin{thebibliography}{99}

\bibitem{RamanProcess}
 A. E. Kozhekin, K. M{\o}lmer, and E. Polzik, ``Quantum memory for light,'' Phys. Rev. A {\bf 62,} 033809/1-5 (2000).
\bibitem{QND}
 B. Julsgaard, J. Sherson, J. I. Cirac, J. Fiur\'{a}\v{s}ek, and E. S. Polzik, ``Experimental demonstration of quantum memory for light,'' Nature (London) {\bf 432,} 482-486 (2004).
\bibitem{DarkSP}
 M. Fleischhauer, and M. D. Lukin, ``Quantum memory for photons: dark-state polaritons,'' Phys. Rev. A {\bf 65,} 022314/1-12 (2002).
\bibitem{EIT}
 D. F. Phillips, A, Fleischhauer, A. Mair, R. L. Walsworth, and M. D. Lukin, ``Storage of light in atomic vapor,'' Phys. Rev. Lett. {\bf 86,} 783-786 (2001).
\bibitem{SinglephotonKuzmich}
 T. Chaneli\'{e}re, D. N. Matsukevich, S. D. Jenkins, S. -Y. Lan, T. A. B. Kennedy, and A. Kuzmich, ``Storage and retrieval of single photons transmitted between remote quantum memories,'' Nature (London) {\bf 438,} 833-836 (2005).
\bibitem{SinglephotonLukin}
 M. D. Eisaman, A. Andr\'{e}, F. Massou, M. Fleischhauer, A. S. Zibrov, and M. D. Lukin, ``Electromagnetically induced transparency with tunable single-photon pulses,'' Nature (London) {\bf 438,} 837-841 (2005).
\bibitem{StorageParametric}
K. Akiba, K. Kashiwagi, T. Yonehara, and M. Kozuma, ``Frequency-filtered storage of parametric fluorescence with electromagnetically induced transparency,'' Phys. Rev. A {\bf 76,} 023812/1-5 (2007).
\bibitem{SSS}
 M. Kitagawa, and M. Ueda, ``Squeezed spin states,'' Phys. Rev. A {\bf 47,} 5138-5143 (1993).
\bibitem{Magnetometry}
 J. Geremia, J. K. Stockton, A. C. Doherty, and H. Mabuchi, ``Quantum Kalman filtering and the Heisenberg limit in atomic magnetometry,'' Phys. Rev. Lett. {\bf 91,} 250801/1-4 (2003).
\bibitem{EITwithSV}
 D. Akamatsu, K. Akiba, and M. Kozuma, ``Electromagnetically induced transparency with squeezed vacuum,'' Phys. Rev. Lett. {\bf 92,} 203602/1-4 (2004).
\bibitem{Slow}
 D. Akamatsu, Y. Yokoi, M. Arikawa, S. Nagatsuka, T. Tanimura, A. Furusawa, and M. Kozuma, ``Ultraslow propagation of squeezed vacuum pulses with electromagnetically induced transparency,'' quant-ph/061109 (to appear in Physical Review Letters).
\bibitem{Odd}
 J. S. Neergaard-Nielsen, B. M. Nielsen, C. Hettich, K. M{\o}lmer, and E. S. Polzik, ``Generation of a superposition of odd photon number states for quantum information networks,'' Phys. Rev. Lett. {\bf 97,} 083604/1-4 (2006).
\bibitem{TimeGated}
 N. Takei, N. Lee, D. Moriyama, J. S. Neergaard-Nielsen, and A. Furusawa, ``Time-gated Einstein-Podolsky-Rosen correlation,'' Phys. Rev. A {\bf 74,} 060101/1-4 (2006).
\bibitem{PPKTP}
 T. Tanimura, D. Akamatsu, Y. Yokoi, A. Furusawa, and M. Kozuma, ``Generation of a squeezed vacuum resonant on a rubidium $D_1$ line with periodically poled KTiOPO$_4$'' Opt. lett. {\bf 31,} 2344-2346 (2006)
\bibitem{FeedFoward}
 M. Kourogi, B. Widiyatmoko, K. Imai, T. Shimizu, and M. Ohtsu, ``Accurate relative frequency cancellation between two independent lasers,'' Opt. Lett. {\bf 24,} 16-18 (1999).
\bibitem{LockFreeze}
 E. S. Polzik, J. Carri, and H. J. Kimble, ``Atomic spectroscopy with squeezed light for sensitivity beyond the vacuum-state limit,'' Appl. Phys. B {\bf 55,} 279-290 (1992).
\bibitem{Quadrature}
 B. Yurke, ``Squeezed-coherent-state generation via four-wave mixers and detection via homodyne detectors,'' Phys. Rev. A {\bf 32,} 300-310 (1985)

\end{thebibliography}
\end{document}